\documentclass{ws-procs9x6}
\newcommand{\be}{\begin{equation}} \newcommand{\ee}{\end{equation}}
\newcommand{\bea}{\begin{eqnarray}} \newcommand{\eea}{\end{eqnarray}}
\newcommand{\bean}{\begin{eqnarray*}} \newcommand{\eean}{\end{eqnarray*}}
\newcommand{\s}[1]{{\scriptscriptstyle #1}}
\newcommand{\sT}{{\s T}}

\begin{document}

\title{Gluonic Pole Matrix Elements in Spectator Models}

\author{ A. Mukherjee$^a$, L. Gamberg$^b$,P. J. Mulders$^c$}

\address{$^a$ Physics Department, Indian Institute of Technology, Powai, 
Mumbai 400076, India.\\
$^b$ Department of Physics, Penn State University-Berks,
Reading, Pennsylvania 19610, USA\\
$^c$ Department of Physics and Astronomy, VU University\\
NL-1081 HV Amsterdam, the Netherlands}

\begin{abstract}
We investigate the gluonic pole matrix element contributing to the first
$p_T$ moment of the distribution and fragmentation functions in a spectator
model. By performing a spectral analysis, we find that for a large class of
spectator models, the contribution of gluonic pole matrix elements is
non-zero for the distribution correlators, whereas in
fragmentation correlators they vanish. This outcome is important in the
study of universality for fragmentation functions.

\end{abstract}

\keywords{Distribution function, fragmentation function, gluonic pole}

\bodymatter

\section{Introduction}

In order to access intrinsic transverse momenta, one needs to do a careful 
 study of the azimuthal dependencies in high energy processes. Azimuthal
imbalance can generate single spin asymmetries. These effects are not
suppressed by powers of hard scale in comparison with the leading order
terms. But it requires measuring hadronic scale quantities (transverse momenta)   
in a high momentum environment. Various symmetries, in particular, time
reversal invariance play a key role here.  

The proper gauge invariant definitions of transverse momentum dependent 
correlators involves path ordered exponentials (gauge links) \cite{collins}.
Recently, it has been found that the part of the path due to the transverse 
gluon field at lightcone $\pm \infty$ are not necessarily suppressed in
light-cone gauge \cite{boer,ji}. It is to be noted that the gauge invariance of
collinear correlators also require a gauge link, however, in this case the
bilocality in the operator is only in the lightcone direction and the gauge
link is universal, independent of the process. Moreover, the link can be set
to unity by choosing an axial gauge. This is not the case for in the
transverse momentum dependent (TMD) correlators, as the bilocality is not
restricted to the lightcone direction but exists also in the transverse direction.
The resulting full color gauge invariant matrix element becomes process
dependent due to the process dependent gauge link. 

As QCD is time reversal invariant, it is possible to distinguish quantities
and observables according to their T-behavior. For distribution correlators
$\Phi(x)$, involving plane wave hadronic states in their definition, 
T-reversal and Hermiticity  implies that the collinear correlators are T-even. 
However the
TMD distribution correlators $\Phi(x,p_T)$  are process dependent due 
to the gauge link as stated above  and time reversal relates $\Phi^{[+]}(x,p_T)$ and         
$\Phi^{[-]}(x,p_T)$, where the link is in two opposite light cone direction. 
Thus one can construct T-even and T-odd combinations
and in general, TMD correlators can be parametrized in terms of both 
T-even and T-odd functions. The T-odd distribution functions, such as Sivers
function, are non-zero due to the presence of the gauge link in the
correlator. 

The situation regarding T-invariance is different for the fragmentation
TMD correlator. These involve hadronic out state in  the definition and thus
one cannot use the T-invariance as a constraint while parametrizing them.
One always has to allow T-even and T-odd parts of the correlator. 
For spin $0$ and spin $1/2$ hadrons, in the collinear case, no T-odd
functions appear at leading twist. Including  TMD, for fragmentation
functions there are now two mechanisms for T-odd terms \cite{boer}:

\begin{itemize}
\item That due to the operator including the gauge link;
\item That due to the final state.
\end{itemize}

A nice feature, however, is that the two mechanisms leading to
T-odd functions can be distinguished. The T-odd operator structure
can be traced back to the color gauge link that necessarily appears in
correlators to render them color gauge-invariant and it is process dependent.
That due to the final state is independent of the process. 

Relating T-odd effects in
different processes, requires azimuthal weighting, which projects
out the transverse momentum weighted parts of the correlators
$\Phi$ and $\Delta$, referred to as (first) transverse moments $\Phi_\partial$
and $\Delta_\partial$, respectively.
 After azimuthal weighting of the cross sections, one simply
finds that the T-odd features originating from the gauge link lead to 
specific factors with which the T-odd functions appear in observables. 
Comparing T-odd effects in distribution functions in semi-inclusive DIS
(SIDIS) and Drell-Yan (DY) processes, one finds a relative minus sign, 
which means that the Sivers function in these two processes differ by a
sign. The T-odd operator parts of  $\Phi_\partial$
and $\Delta_\partial$ are precisely the soft limits ($k_1\rightarrow 0$) 
of the gluonic pole matrix elements~\cite{boer}, denoted by $\phi_G(k_1=0)$
and $\Delta_G(k_1=0)$. In this work, we investigate the spectral properties 
of these matrix elements, which play an important role in the universality
issue of the TMD distribution and fragmentation functions.  

\begin{figure}
\begin{center}
\includegraphics[width=4cm]{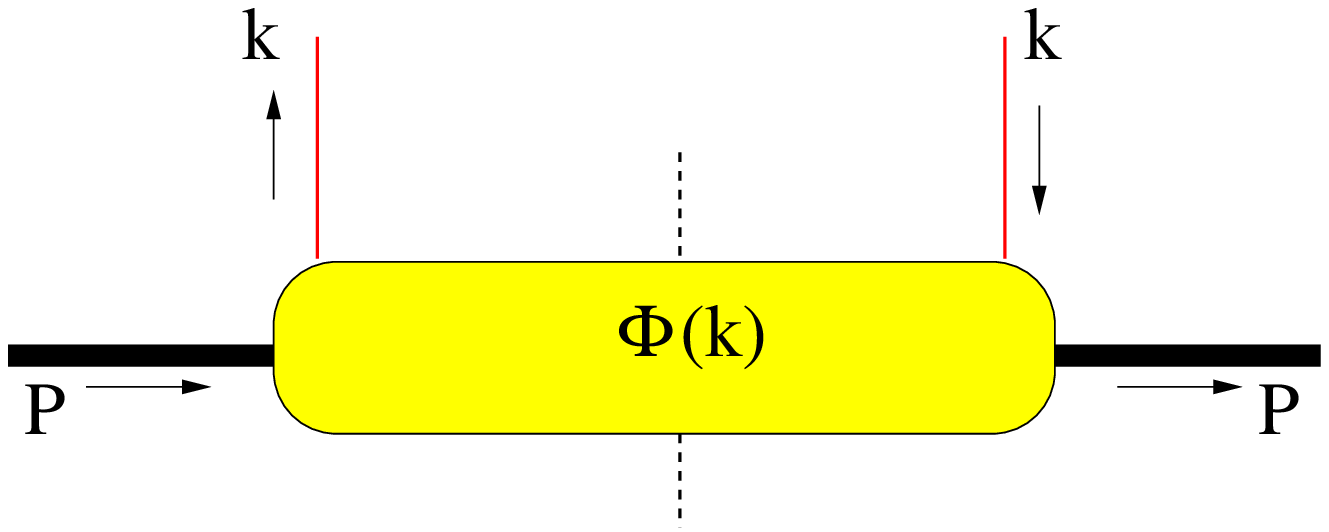}
\hspace{2cm}
\includegraphics[width=4cm]{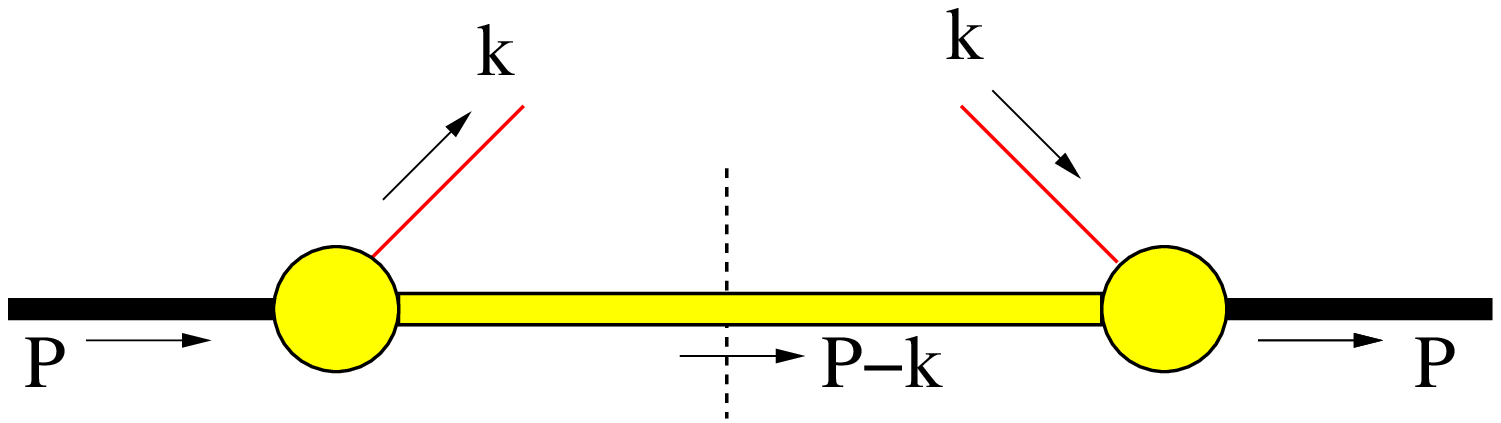}
\\
(a)\hspace{6cm}(b)
\end{center}
\caption{\label{qqspec-1}
The graphical representation of the correlator in the case of
distributions of partons with momentum $k$ in a hadron with
momentum $P$ and the spectator model description.}
\end{figure}
\begin{figure}
\begin{center}
\mbox{}\hspace{-0.6cm}\includegraphics[width=3cm]{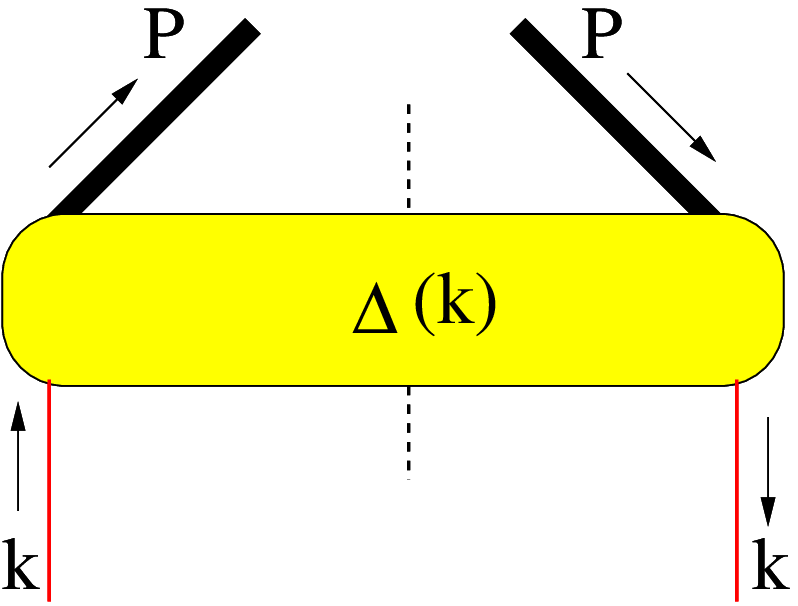}
\hspace{2cm}
\includegraphics[width=4cm]{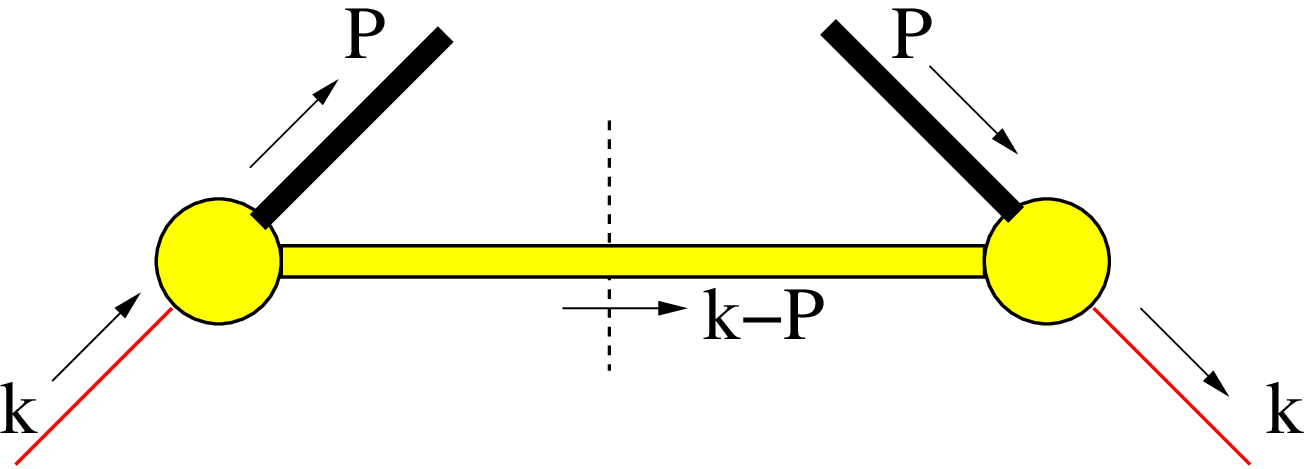}
\\
(a)\hspace{6cm}(b)
\end{center}
\caption{\label{qqspec-2}
The graphical representation of the correlator in the case of
fragmentation of partons with momentum $k$ into a hadron with
momentum $P$ (a) and the spectator model description (b).}
\end{figure}

For the correlators, depicted in Figs~\ref{qqspec-1}
and \ref{qqspec-2}, the expressions in terms of bilocal matrix
elements are frequently used as a  starting point in modeling
distribution and fragmentation functions.
In particular the spectator model has become
fairly popular, because it is easy, flexible and intuitively
attractive. On the other hand, one should be very careful, because
the predictive power depends on limiting oneself in the choice of
spectator (e.g.\ a diquark with fixed mass for the nucleon) and
using simple vertices. In fact making a spectral analysis of
the spectator and allowing for the most general vertices one
would lose all predictive power. Here
we will investigate differences between distribution
and fragmentation functions using a spectral analysis and using
physical intuition in restricting the momentum dependence and
asymptotic behavior of the vertices.
         
\begin{figure}
\begin{center}
\includegraphics[width=5cm]{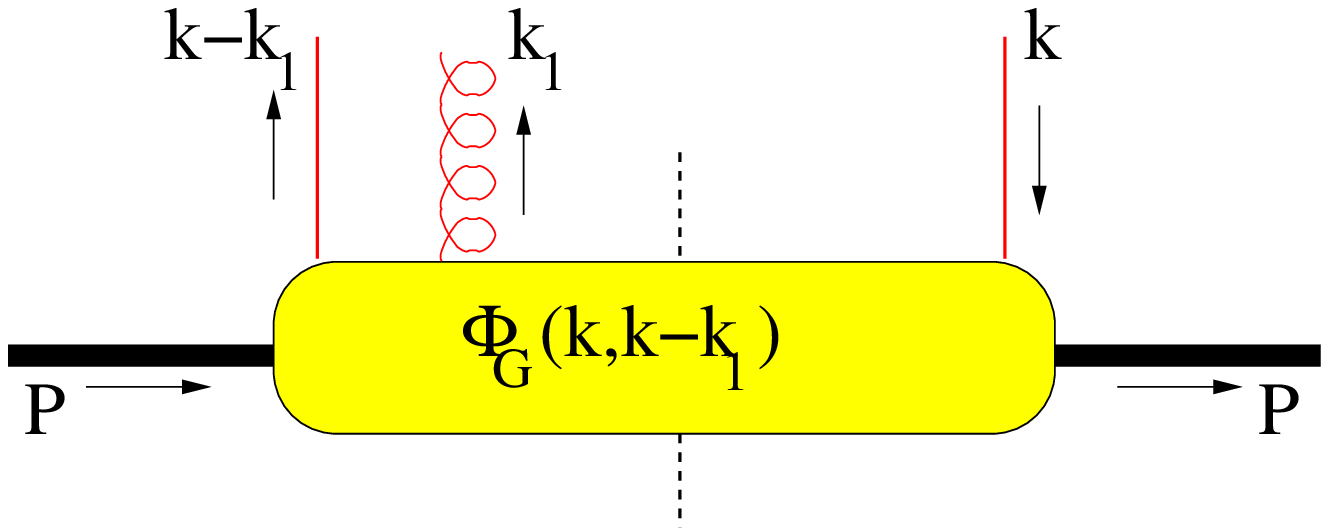}
\\
\includegraphics[width=4.5cm]{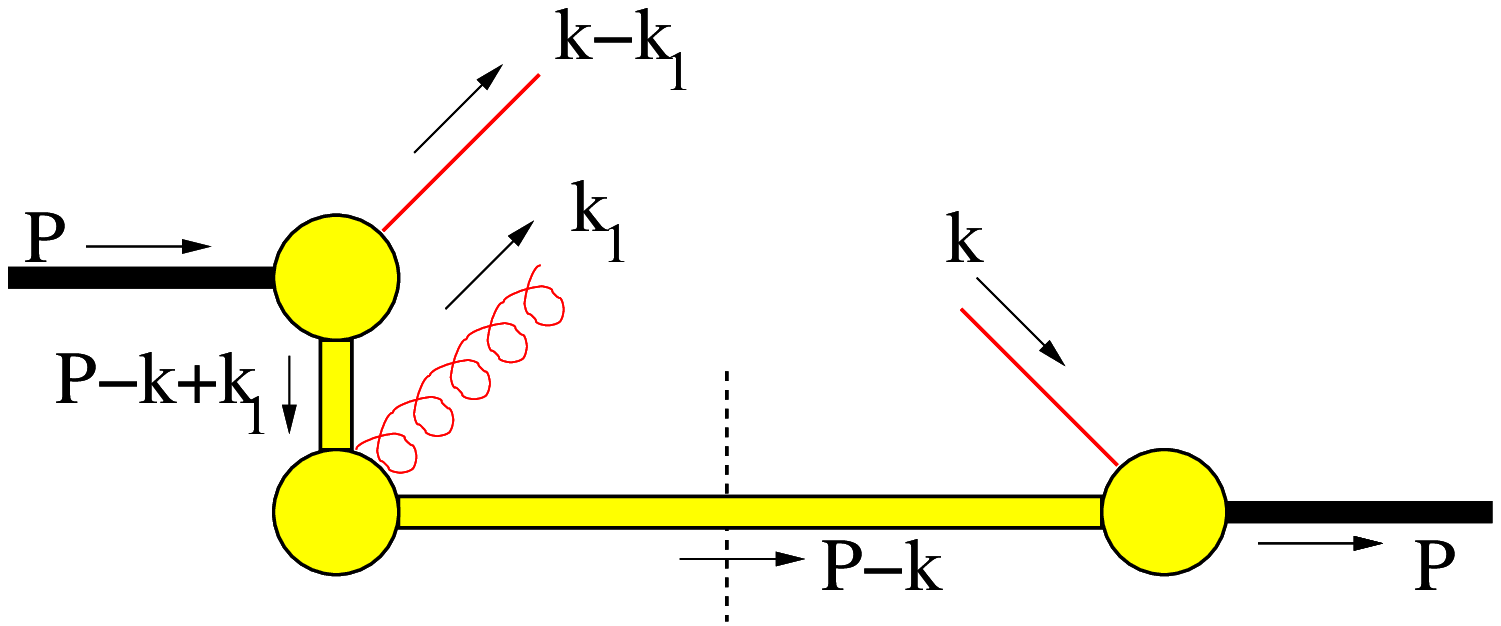}
\mbox{}\hspace{2cm}
\includegraphics[width=4.5cm]{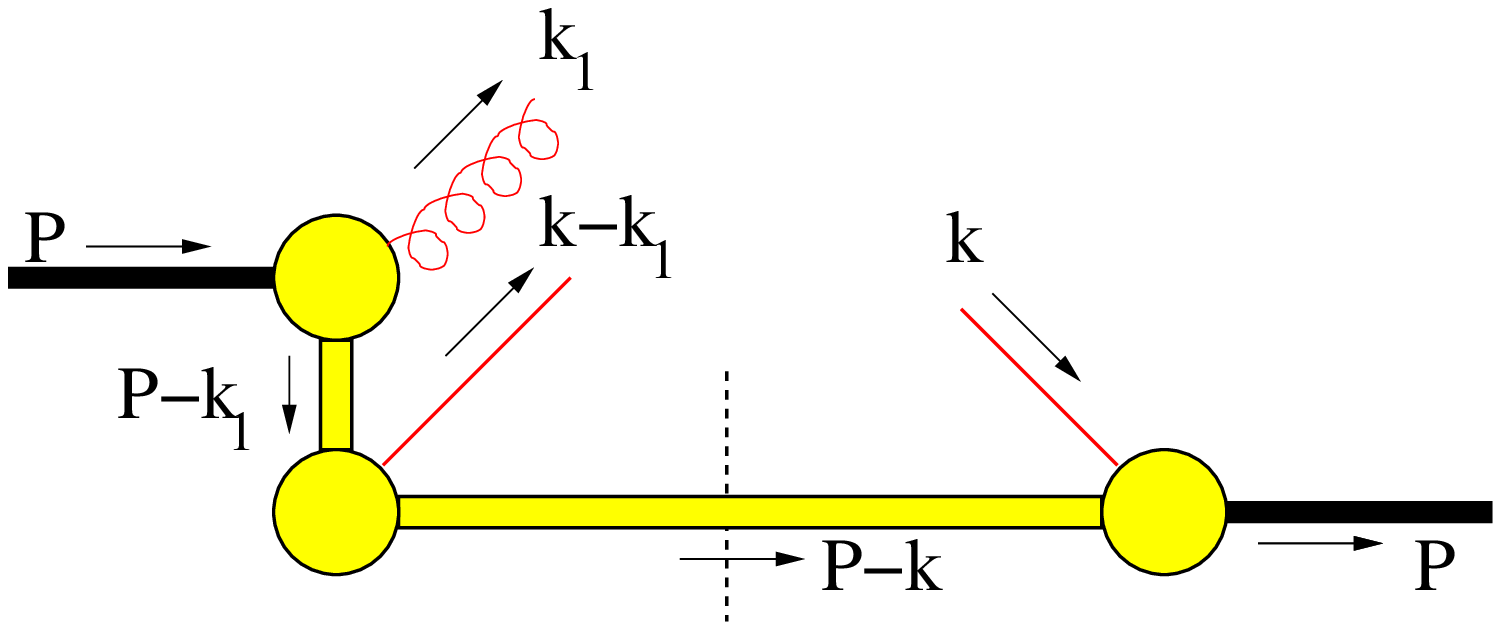}
\\ \mbox{}(a)\hspace{6cm}(b)
\end{center}
\caption{\label{qqGspec-1}
The graphical representation of the quark-quark-gluon
correlator $\Phi_G$ in the case of distributions including
besides the quark a gluon with momentum $k_1$ and the
possible intermediate states (a) and (b) in a spectator
model description.}
\end{figure}

\begin{figure}
\begin{center}
\includegraphics[width=4cm]{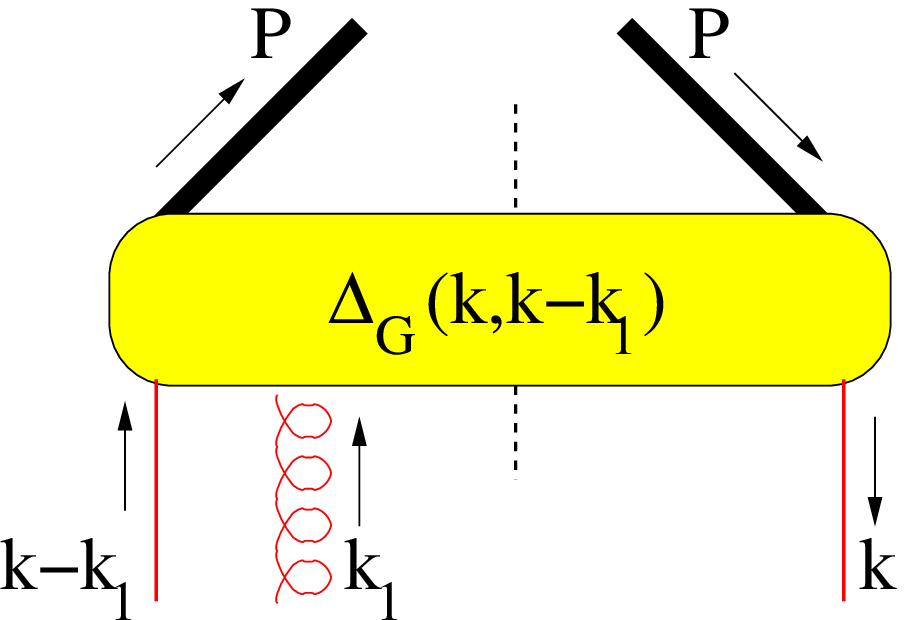}
\\
\includegraphics[width=4.5cm]{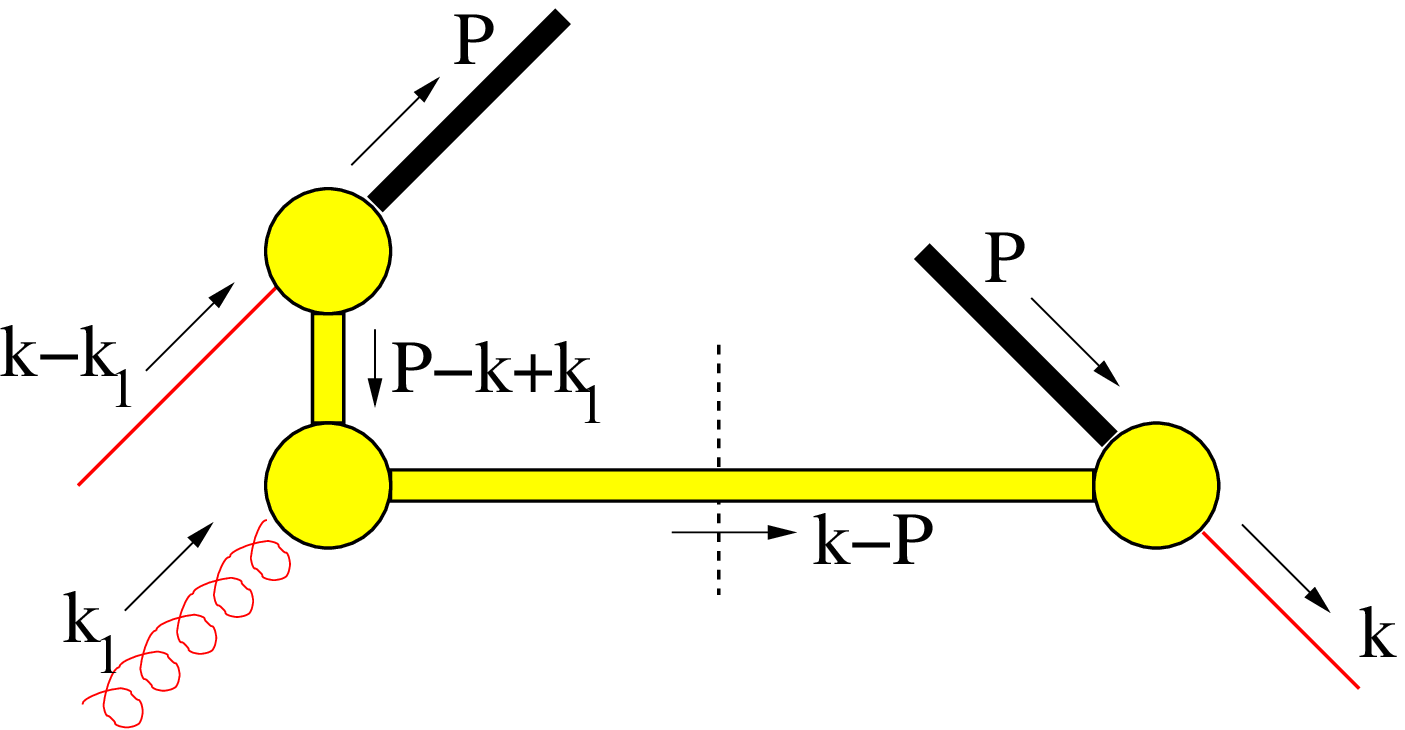}
\mbox{}\hspace{2cm}
\includegraphics[width=4.5cm]{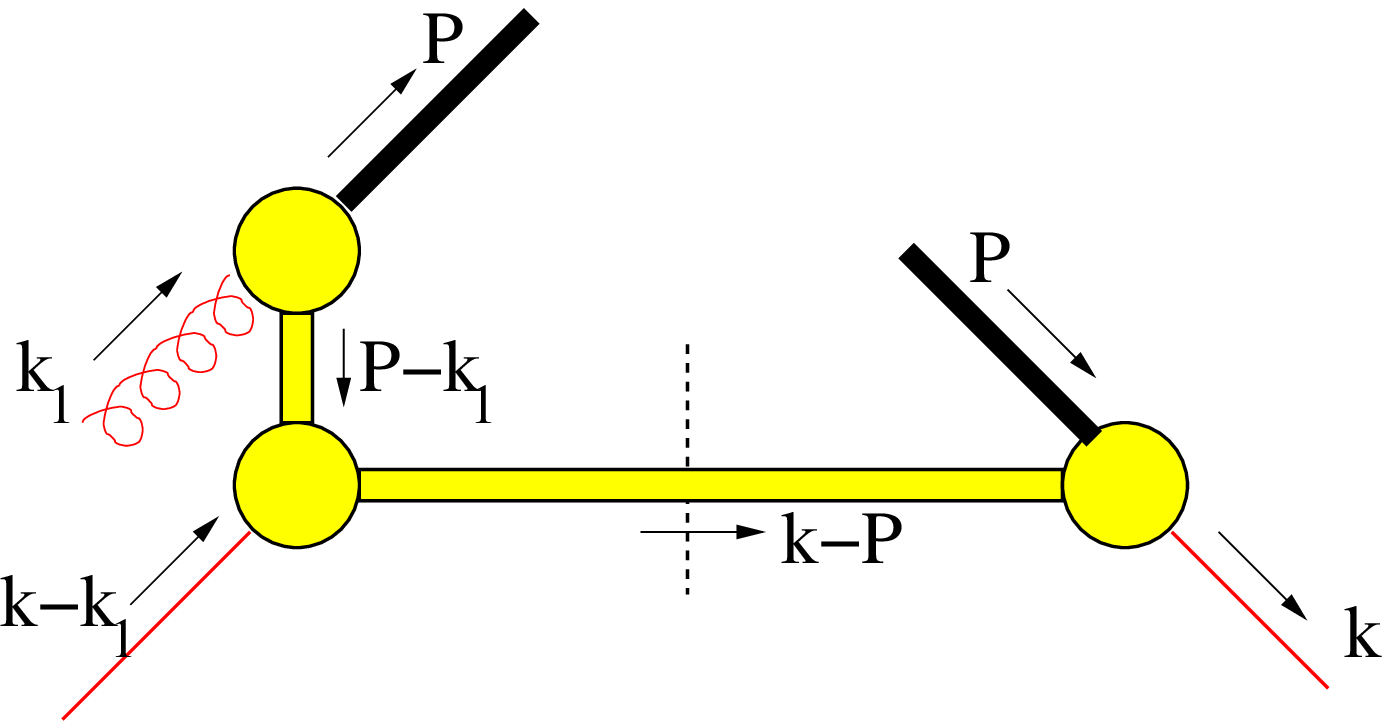}
\\ \mbox{}(a)\hspace{6cm}(b)
\end{center}
\caption{\label{qqGspec-2}
The graphical representation of the quark-quark-gluon
correlator $\Delta_G$ in the case of fragmentation including
besides the quark a gluon with momentum $k_1$ and the
possible intermediate states (a) and (b) in a spectator
model description.}
\end{figure}

Rather than looking at the full process in the model and 
carefully studying the cuts \cite{metz,collins1}, we look 
at the soft part only \cite{whepp}, but do this for the quark-quark-gluon
correlators $\Phi_G$ and $\Delta_G$, as shown in Figs \ref{qqGspec-1}
and \ref{qqGspec-2}, respectively. Our approach starting directly 
with the
multi-parton correlator has the advantage that we can work at tree-level
and just do a spectral analysis, unlike a one loop calculation
\cite{gamberg}.

\section{Gluonic pole matrix elements\label{GPsection}}

We make a Sudakov decomposition
of the momenta of active partons, $k=x\,P+\sigma\,n+k_\sT$. The
Sudakov vector $n$ is an arbitrary light-like four-vector $n^2=0$
that has non-zero overlap $P\cdot n$ with the hadron's momentum $P$. We
will simply choose $P\cdot n = 1$.

The TMD distribution correlators are given by :
\begin{equation}\label{TMDDF}
\Phi_{ij}^{[\mathcal U]}(x{,}k_\sT)
={\int}\frac{d(\xi{\cdot}P)\,d^2\xi_\sT}{(2\pi)^3}\ e^{ik\cdot\xi}\,
\langle P|\,\overline\psi_j(0)\,\mathcal U_{[0;\xi]}\,
\psi_i(\xi)\,|P\rangle\big\rfloor_{\text{LF}}\ .
\end{equation}
The \emph{Wilson line} or \emph{gauge link}
$\mathcal U_{[\eta;\xi]}
=\mathcal P{\exp}\big[{-}ig{\int_C}\,ds{\cdot}A^a(s)\,t^a\,\big]$
is a path-ordered exponential along the integration path $C$ with
endpoints at $\eta$ and $\xi$.

In azimuthal asymmetries one needs the transverse moments contained in the
correlator
\be
\label{TransverseMoment}
\Phi_{\partial}^{\alpha\,[\mathcal U]}(x)
= \int d^2k_\sT\ k_\sT^\alpha\,\Phi^{[\mathcal U]}(x{,}k_\sT)\, .
\ee

This can be written as 
\begin{equation}\label{DECOMPOSITION}
\Phi_\partial^{\alpha\,[\mathcal U]}(x)
=\tilde \Phi_\partial^{\alpha}(x)
+C_G^{[\mathcal U]}\,\pi\Phi_G^{\alpha}(x,x) ,
\end{equation}
with calculable process-dependent gluonic pole factors $C_G^{[\mathcal U]}$
and process (link) independent correlators $\tilde\Phi_\partial$ (T even) 
and $\Phi_G$ (T-odd).  
The latter is precisely the soft limit $x_1\rightarrow 0$ of
a quark-gluon correlator $\Phi_G(x,x_1)$ of the type
\bea
\label{GP}
\Phi_G^\alpha(x,x{-}x_1)
 = n_\mu
\int\frac{d(\xi{\cdot}P)}{2\pi}\frac{d(\eta{\cdot}P)}{2\pi}\
e^{ix_1(\eta\cdot P)}e^{i(x-x_1)(\xi\cdot P)}\nonumber\\
\langle P|\,\overline\psi(0)\,U_{[0;\eta]}^n\,gG^{\mu\alpha}(\eta)\,
U_{[\eta;\xi]}^n\,\psi(\xi)\,|P\rangle\,\big\rfloor_{\text{LC}}\ .
\eea
The TMD fragmentation
correlator is given by~\cite{boer}
\bea
\Delta^{[\mathcal U]}_{ij}(z,k_\sT)&=&
\sum_X\int\frac{d(\xi\cdot P_h)\,d^2\xi_\sT}{(2\pi)^3} e^{i\,k\cdot\xi}
\langle 0 |\mathcal U_{[0,\xi]}\psi_i(\xi)|P,X\rangle
\nonumber\\&&~~~~~~~~~~\langle P,X|\bar{\psi}_j(0)|0\rangle |_{LF}\,  .
\label{TMDFF}
\eea
In the transverse moments obtained after $k_\sT$-weighting,
\bea
\Delta^{\alpha\,[\mathcal U]}_{\partial}(z)
=\int d^2k_\sT\ k_\sT^\alpha \Delta^{[\mathcal U]}(z,k_\sT)
=\tilde{\Delta}_\partial^\alpha\left(\tfrac{1}{z}\right)
+C_G^{[\mathcal U]}
\,\pi\Delta_G^\alpha\left(\tfrac{1}{z},\tfrac{1}{z}\right);
\label{decompfrag}
\eea
the two link independent correlators $\tilde\Delta_\partial$
and $\Delta_G$ contain again a T-even and T-odd operator combination,
respectively, 
however, parametrizations of  both of them contain T-odd functions 
as the final state is involved in both. The gluonic pole correlator is the soft limit,
$z_1^{-1} = x_1 \rightarrow 0$, of the quark-gluon correlator
\bea
\Delta_{G\,ij}^\alpha\left(x,x-x_1\right)
&=&\sum_X \int\frac{d(\xi{\cdot}P)}{2\pi}\frac{d(\eta{\cdot}P)}{2\pi}\
e^{i\,x_1(\eta\cdot P)}e^{i\,(x-x_1)(\xi\cdot P)}\,
\nonumber  \\&&\times
\langle 0 | \mathcal U^n_{[0,\eta]}\, gG^{n\alpha}(\eta)
\,\mathcal U^n_{[\eta,\xi]}\psi_i(\xi)|P,X\rangle
\langle P,X|\overline{\psi}_j(0)|0\rangle\Bigg|_{LC} .
\label{GLa}
\eea
With $\Delta_G(x,x)$ being zero, T-odd FFs in $\Delta_\partial$
only come from the T-even operator combination in $\tilde \Delta_\partial$,
which is due to the hadron-jet state and they are process independent,
for instance the T-odd Collins function. In contrast T-odd DFs in
$\Phi_\partial$ only can come from $\Phi_G(x,x)$. They can still
be universal but acquire process dependent gluonic pole
factors~\cite{bomhof,bomhof1}. We calculate $\Phi_G$ and $\Delta_G$ in a
spectator model approach.

\section{The spectator model approach}

In a typical spectator model approach to distribution or fragmentation
correlators one considers a spectator with mass $M_s$. The
result for the cut, but untruncated, diagrams, such as in
Figs.~\ref{qqspec-1} and \ref{qqspec-2} are of the form
\be
\Phi(x,k_\sT) \sim \int d(k\cdot P)
\ \frac{F(k^2,k\cdot P)}{(k^2-m^2+i\epsilon)^2}
\,\delta\left( (k-P)^2 - M_s^2\right).
\label{basic}
\ee
The details of the numerator function depend on the details of the
model, including the vertices, polarization sums, etc. These must
be chosen in such a way as to not produce unphysical effects,
such as a decaying proton if $M \ge m+M_s$. Thus $m$ in Eq.~\ref{basic}
must represent some constituent mass in the quark propagator, rather
than the bare mass.
The useful feature of the model is its ability to
produce reasonable valence and even sea quark distributions using the
freedom in the model connected to an intuitive picture.
The results for the fragmentation function in the spectator
model is identical upon the substitution of $x = 1/z$.

The quark-gluon correlators  as shown in Figs. ~\ref{qqGspec-1} and
\ref{qqGspec-2} can be calculated in the spectator model and the gluonic
pole matrix elements can be extracted by taking the limit $x_1 \rightarrow
0$. 
Assuming that the numerator does not grow with $k_1^-$ one can
easily perform the $k_1^-$ integrations \cite{whepp}.
We obtain (for $x \ge 0$)
\bea
\Phi_G(x,x)
&=& - \int d^2k_\sT\,d^2k_{1\sT}
\ \frac{(1-x)\,F_1(x,0,k_\sT,k_{1\sT})\theta(1-x)}{
\bigl(\mu^2-k_\sT^2\bigr) \bigl(x\,B_1+(1-x)\,A_2\bigr)\,A_1}\, .
\eea
and for fragmentation functions ($x = 1/z \ge 1$)
\bea
\Delta_G(x,x) &=& 0\, .
\eea
\section{Conclusion}
We have investigated the gluonic pole contributions to the
distribution and fragmentation functions. Instead of doing a
quantitative analysis involving details of a phenomenological model, we
limit ourselves to a spectral analysis within the spectator framework,
in order to understand the basic
features of these quantities. 
We simply assumed that masses and vertices do not spoil our
analysis.
We find that under realistic assumptions, the gluonic pole
contributions for  fragmentation correlators vanish whereas
these contributions do not vanish for distribution correlators.
The result for fragmentation correlators at nonzero gluon momentum
is nonzero. A full proof that the gluonic pole contributions to the
fragmentation correlators vanish  is important as it eliminates a whole class of matrix
elements parameterized in terms of T-odd fragmentation functions besides
the T-odd fragmentation functions
in the parameterization of the two-parton correlators.
\section{Acknowledgments}
AM would like to thank the organizers of Transversity 2008 for the kind
invitation and support.


\end{document}